\documentclass[aps,showpacs, nofootinbib]{revtex4}
\newcommand{\be}{\begin{equation}}
\newcommand{\ee}{\end{equation}}
\newcommand{\ben}{\begin{eqnarray}}
\newcommand{\een}{\end{eqnarray}}

\newcommand{\cO}{{\cal O}}

\newcommand{\p}{\partial}
\newcommand{\na}{\nabla}

\newcommand{\tg}{\tilde g}

\newcommand{\tR}{\tilde R}
\newcommand{\tW}{\tilde W}

\newcommand{\teta}{\tilde \eta}


\begin{document}

\title{Matter Outside a Static Higher-Dimensional Black Hole}
%%%%%%%%%%%%%%%%%%%%%%%%%%%%%%%%%%%%%%%%%%%%%%%%%%%%%%%%%%%%%%

\author{Marek Rogatko}
\affiliation{Institute of Physics \protect \\
Maria Curie-Sklodowska University \protect \\
20-031 Lublin, pl.~Marii Curie-Sklodowskiej 1, Poland \protect \\
marek.rogat@poczta.umcs.lublin.pl \protect \\
rogat@kft.umcs.lublin.pl}

%%%%%%%%%%%%%%%%%%%%%%%%%%%%%%%%%%%%%%%%%%%%%%%%%%%%%%%%%%%%%%%%%%%%
\date{\today}
%\pacs{04.30.Nk, 04.40.-b}

%%%%%%%%%%%%%%%%%%%%%%%%%%%%%%%%%%%%%%%%%%%%%%%%%%%%%%%%%%%%%%%%%%%%%%%%%%%%%%%%%%%%%%%%%%%%%%%%%%%
\begin{abstract}
We considered matter fields composed of a perfect fluid in the static 
higher-dimensional spherically symmetric 
asymptotically flat black hole spacetime. The proof of the nonexistence of perfect fluid matter 
in such background was provided under the auxiliary condition, which can be interpreted as a relation
connecting stellar mass and black hole mass in question.
 
\end{abstract}
%%%%%%%%%%%%%%%%%%%%%%%%%%%%%%%%%%%%%%%%%%%%%%%%%%%%%%%%%%%%%%%%%%%%%%%%%%%%%%%%%%%%%%%%%%%%%%%%%%%%

\maketitle

%%%%%%%%%%%%%%%%%%%%%%%%%%%%%%%%%%%%%%%%%%%%%%%%%%%%%%%%%%%%%%%%%%%%%%%%%%%%%%%%%%%%%%%%%%%%%%%%%%%%%
\section{Introduction}
One of the most important issues of gravity theory are connected with a gravitational collapse
and the emergence of black holes. In the case of general relativity the uniqueness theorem for black holes
states that the static electrovac black hole solution is diffeomorphic to the domain of outer communication
of Reissner-Nordstr\"om (RN) spacetime whereas the rotating one is diffeomorphic to Kerr-Newman (KN)
spacetime. The first classification of non-singular black hole solutions was presented in Refs.
\cite{isr67}-\cite{heu94}. The condition of non-degeneracy of the event horizon in the proofs of
the uniqueness theorem was eliminated in \cite{chr99a,chr99b}. In the case of static electro-vacuum
black holes with degenerate components of the event horizons, the study of near-horizon geometry
enables to finish the classification \cite{chr07}.
\par
The problem of stationary axisymmetric black hole solutions was far more complicated. It was treated
in Refs.\cite{car73}-\cite{rob75}, while the complete proof was found by Mazur \cite{maz82} and Bunting
\cite{bun}-\cite{car85} (for a review of the uniqueness of black hole solutions see, e.g., 
\cite{heub} and references therein).
\par
It turned out that M/string theory attempts of unifications of all known forces of Nature
also triggered works bounded with the mathematical aspects of black holes in the low-energy 
string theory as well as its higher dimensional generalization. Various aspects of the low-energy
string black holes including the staticity theorem, uniqueness theorems in Einstein-Maxwell 
axion dilaton (EMAD) gravity, dilaton gravity with auxiliary $U(1)$-gauge fields and supergravity
theories were widely treated \cite{lowen}.\\
The uniqueness theorem for static $n$-dimensional black hole, both in vacuum and
charged case is well established \cite{gib}. The complete 
classification of $n$-dimensional
charged black holes having both degenerate and non-degenerate
components of event horizon was presented in Refs.\cite{rog0306}.
In Ref.\cite{rog04}, taking into account the both {\it electric} and {\it magnetic} components of 
$(n-2)$-gauge form $F_{\mu_{1} \dots \mu_{n-2}}$ the uniqueness
of static higher dimensional {\it electrically} and {\it magnetically} charged
black hole containing an asymptotically flat hypersurface with compact interior
and non-degenerate components of the event horizon was proved. On the other hand, the staticity theorem
for generalized Einstein-Maxwell (EM) system was discussed in \cite{rog05}.   
\par
Uniqueness theorem for stationary axisymmetric
$n$-dimensional black holes is much more delicate problem.
It turns out that generalization of Kerr metric to arbitrary $n$-dimensions proposed by
Myers-Perry \cite{mye86} is not unique. 
Rotating black hole ring solution with the same angular momentum and mass
but the horizon homeomorphic to $S^{2} \times S^{1}$ presented in
\cite{emp02} constitutes the counterexample. However, the existence of black rings is consistent 
with the generalization of Hawking's theorem \cite{haw73}. Namely, as was shown in 
\cite{gal05,hel05} cross sections of event horizons and
outer horizons were of the positive Yamabe type. They admit metrics of positive scalar curvature.
\par
Recent results concerning the attempts of treating the problem of uniqueness of stationary
axisymmetric higher dimensional black objects (black holes and black rings) were presented in
Refs.\cite{nrot}.
\par
The other tantalizing question in mathematical
theory of black holes is 
concerned with the uniqueness problem for stationary or static black holes and matter.
Depending on the matter model in question, black holes 
may allow to exist the nontrivial fields outside its event horizon.
This question was elaborated in Refs.\cite{haj74}.
It was also revealed \cite{loh84}
RN black hole solution with both an electric and magnetic charges can be destroyed in the presence
of a massless Dirac fermion field. On the other hand, it was shown \cite{smo} that the only
black hole solutions of four-dimensional spinor Einstein-dilaton-Yang-Mills field equations
of motion were those for which spinors vanished identically outside black hole. It means that Dirac fermion
fields either enter the black hole in question or escape to infinity.
This tendency was also confirmed in Ref.\cite{nak12}.
\par
On the other hand, the proof that the stellar model composed of perfect fluid, nonrotating, 
self-gravitating and physically isolated is spherical symmetric has also a long story.
Significant progress was made in Refs.\cite{mas87}-\cite{bei92}. Among all, it was shown that
the spatial part of the underlying geometry of static stellar model ought to be conformally flat,
due to the consequence of the positive mass theorem. Because of the fact that spatial conformal
flatness is equivalent to the spherical symmetry in static stellar model \cite{lin80},
the above arguments completed the proof. However, in Ref.\cite{lin94} various technical assumptions
were get rid of and unphysical restrictions on the equation of state of perfect fluid were weaken.
Recently, it was shown \cite{shi06}
that matter configuration composed of
a perfect fluid could not be at rest outside a four-dimensional black hole in asymptotically flat 
static spacetime.
\par
The main motivation for the present work stems from the fact that many contemporary
theories of unifications of the all known forces in Nature like superstring/M-theory require
extra dimensions to be consistently formulated. The theoretical properties of the higher-dimensional
gravitational field could be quite different. The appearance of black rings, black Saturns and black lens
objects confirms this conjecture. So it will be of much interest to get an insight into the
multidimensional physics which may lead to a wider understanding of the gravitational field.
Just, motivated by higher dimensional gravity theories which are crucial ingredients
in modern unification schemes,
we shall study the problem of the uniqueness of static $n$-dimensional black hole and a perfect fluid
matter outside the black hole event horizon. We provide a proof for the nonexistence of
a perfect fluid star outside the black hole event horizon, under the condition that star mass is smaller 
comparing to the black hole one.
\par
Our paper is organized as follows. In Sec.II we describe higher dimensional Einstein perfect fluid system
and introduce the quantities which are needed in the proof. In Sec.III we introduce the conformal
transformation of the spatial part of the line element in question and establish important properties of it.
Sec.IV will be devoted to the proof of the nonexistence of a 
perfect fluid matter in static, $n$-dimensional black hole background. In Sec.V we provide the physical
meaning of the inequality which is crucial in the uniqueness proof. Then, one concludes the investigations.

%%%%%%%%%%%%%%%%%
%%%%%%%%%%%%%%%%%%%%%%%%%%%%%%%%%%%%%%%%%%%%%%%%%%%%%%%%%%%%%%%%%%%%%%%%%%%%%%%%%%%%%%%%%%%%%%%%
\section{Higher dimensional Einstein Perfect Fluid System}
In this section we shall consider higher dimensional Einstein field equations with perfect fluid matter
which is in static equilibrium. A static spacetime admits a hypersurface orthogonal to asymptotically
timelike Killing vector field $k_\alpha$. Suppose that $t$ will be a function which level surface
is orthogonal to $k_\alpha$ and $t^\beta~\p_\beta t = 1$. Thus, the line element 
of $n$-dimensional static spacetime with the asymptotically timelike Killing vector field 
$k_{\alpha} = \big({\p \over \p t }\big)_{\alpha}$
and $V^{2} = - k_{\mu}k^{\mu}$ is subject to the relation
\be
ds^2 = - V^2 dt^2 + g_{i j}dx^{i}dx^{j},
\ee
where $g_{i j}$ is an induced metric on 
the hypersurface orthogonal to $k_{\alpha}$, i.e., the metric on
$t = const$ submanifold. $V$ and $g_{i j}$
are independent of the $t$-coordinate as the quantities
of the hypersurface $\Sigma$ of constant $t$. 
\par
The Einstein perfect fluid equations of motion are provided by 
\ben \label{m1}
D_a~D^a V &=&  {8~\pi~V \over n - 2}~\bigg[
(n - 3)~\rho + (n - 1)~p \bigg], \\ \label{m2}
{}^{(g)}R_{ij} &=& {8~\pi~g_{ij} \over n - 2}~\bigg( \rho - p \bigg) + {1 \over V}~D_i~D_j V,\\ \label{m3}
D_c p &=& - {D_c V \over V}~\bigg( \rho + p \bigg),
\een 
where $D_i$ and ${}^{(g)}R_{ij}$ denote covariant derivative and Ricci curvature tensor ${}^{(g)}R_{ij}$ on
$(\Sigma,~g_{ij})$, respectively.\\
In addition there is a perfect fluid matter characterized by the energy density $\rho$ and 
pressure $p$ in a spacetime in question. The perfect fluid fulfills an equation of state of the form
$p = p(\rho)$. On the other hand, the surface of a perfect fluid star is a $(n-2)$-dimensional closed,
connected equipotential surface for which $V(x) = V_S > 0$.\\
As in Ref.\cite{mye86}
we define mass of the higher dimensional object as
\be M = {(n-2)~\Omega_{n-2} \over 8~\pi}~m_\mu,
\ee
where $\Omega_{n-2}$ is the volume of $(n-2)$-dimensional unit sphere. On this account, we have
\be
M = \int_{0}^{r} {2~\pi^{{n-1 \over 2}} \over \Gamma ({n-1 \over 2})}~r^{n-2}_\mu~\rho~dr = 
{(n-2)~\Omega_{n-2} \over 8~\pi}~m_\mu.
\ee
According to the above relation one gets that
\be
m_\mu = {8~\pi \over (n-1)~(n-2)}~\rho~r^{n-1}_\mu.
\ee
In the asymptotic flat spacetime we can find an appropriate coordinate system which provides
the following asymptotical behaviour of $V$:
\be
V = 1 - {m_\mu \over r^{n - 3}} + {\cal O}\bigg( {1 \over  r^{n - 2}} \bigg),
\label{asv}
\ee
and accordingly  for the metric tensor we obtain the relation
\be
g_{ij} = \bigg( 1 + {2 ~m_\mu \over r^{n - 3}} \bigg) \delta_{ij} +
{\cal O} \bigg( {1 \over r^{n - 2}} \bigg).
\ee
As was pointed out in four dimensional treatment of this subject, equation (\ref{m3}) provides the fact
that the surface $\rho = const$ is identical with the surface $V = const$. 
Moreover, this relation indicates
that $D_i p$ diverges at the black hole event horizon, where one has that $V = 0$.
In the case of a black hole, the hypersurface $\Sigma$ is a simply connected spacelike
hypersurface and $\bar \Sigma$ denotes the closure of it. The topological
boundary of $\Sigma$, ~$\p \Sigma = \bar \Sigma \backslash \Sigma$ is a nonempty
topological manifold with $g_{ij}k^{i}k^{j} = 0$ on $\p \Sigma$, which constitutes
black hole event horizon. In a static spacetime it coincides with the Killing horizon.\\
Due to the fact of $D_i p$ divergence on $V = 0$, which constitutes 
an unphysical situation, we shall consider the case where stellar surface will be disjoint
from black hole event horizon. \\
The other argument about disjointness of the star from the black hole event horizon can be provided by the theory
of elliptic differential equations. 
Because of the fact that equation (\ref{m1}) is of elliptic type, the standard boundary elliptic 
estimate provides that near the event horizon,
the norm of the gradient of $V$ is bounded by the norm
of the right-hand side of equation (\ref{m1}) and that of $V$ \cite{gil01}. The surface of the star is the level set
$V = V_S > 0$, while the horizon is zero set of $V$. Just the gradient bound for $V$ gives a positive lower 
bound for the distance between star and horizon level sets. This in turn yields that the surface of the star
is disjoint from the black hole event horizon.
\par
%%%%%%%%%%%%%%%%%%%%%%%%%%%%%%%%%%%%%%%%%%%%%%%%%%%%%%%%%%%%%%%%%%%%%%%%%%%%%%%%%%%%%%%%%%%%
Because of the fact that $V = const$ surface is identical with $\rho = const$ surface and having in mind
equation of state where $p$ is function of density, one can regard both pressure and density as functions
of $V$. The minimum value of $V$ takes place inside the considered star.
To proceed further we can define functions $r_\mu(V)$ and $m_\mu(V)$, which
are solutions of the following differential equations:
\ben \label{rv}
{d r_{\mu} \over d V} &=& {r_{\mu} \bigg( r^{n-3}_{\mu} - 2~m_\mu \bigg) \over
V(r)~\bigg[ {8 \pi \over n - 2}~
p~{r^{n-1}_{\mu}} + m_\mu~(n-3) \bigg]}, \\ \label{mv}
{d m_{\mu} \over d V} &=&
{8~\pi \over (n-2) }~
{\rho~{r^{n-1}_{\mu}}~\bigg( r^{n-3}_{\mu} - 2~\mu \bigg) \over 
V(r)~\bigg[ {8 \pi \over n - 2}~
p~{r^{n-1}_{\mu}} + m_\mu~(n-3) \bigg]}.
\een
The above differential equations satisfy the boundary conditions
\ben
r_\mu(V_S) &=& R_\mu = \bigg(
{2~\mu \over 1 -V^2_S} \bigg)^{1 \over n - 3},\\
m_\mu(V_S) &=& \mu,
\een 
where $\mu$ is a constant value which can be interpreted as a local mass of the 
perfect fluid star in question.\\
Let us introduce on the domain where the solutions of equations (\ref{rv}) and (\ref{mv}) exist,
the function $W_{\mu}(V)$ provided by
the relation of the form
\be
W_{\mu} = \bigg( 1 - {2~m_\mu \over r^{n - 3}_\mu} \bigg)~\bigg( {d r_{\mu} \over d V} \bigg)^{-2}
= {V^2~\bigg[ {8 \pi \over n - 2}~
p~{r^{n-1}_{\mu}} + m_\mu~(n-3) \bigg]^2 \over
{r^{n-1}_{\mu}}~\bigg( {r^{n-3}_{\mu}} - 2~m_\mu \bigg)}.
\label{wdef}
\ee
The above definition is valid inside the considered star. Consequently, outside the star it is given by the expression
\be
W_{\mu} = {
(n-3)^2~\bigg( 1 -V^2 \bigg)^{2 \bigg( {n-2 \over n-3} \bigg)} \over
2^{2n-4 \over n-3}~\mu^{2 \over n-3}}.
\ee
Hence, having in mind equation (\ref{asv}), the asymptotic behaviour of $W_\mu$
near spatial infinity is subject to the relation
\be
W_\mu = {(n-3)^2 \over \mu^{2 \over n-3}}~\bigg[
{m^{2~( {n-2 \over n-3})}_\mu \over r^{2(n-2)}} \bigg] + \cO \bigg(
{1 \over r^{2n -3}} \bigg).
\ee
At the surface of the star one has the following conditions:
\ben
r_\mu = R_\mu > \bigg( 2~\mu \bigg)^{1 \over n-3} &=& \bigg( 2~m_\mu \bigg)^{1 \over n-3} > \\ \nonumber
\bigg[ &-& {16~\pi ~p ~r^{n-1}_\mu \over (n-3)~(n-2)} \bigg]^{1 \over n-3}= 0.
\een
By virtue of the above, the derivatives ${d r_{\mu} \over d V}$ and ${d m_{\mu} \over d V}$ yield
\ben
{d r_{\mu} \over d V} \mid_{V = V_S} &=& {R^{n-2}_{\mu}~V_S \over m_\mu~(n-3)},\\
{d m_{\mu} \over d V} \mid_{V = V_S} &=& {8~\pi \over (n-2) }~
{\rho~R^{2n-4}_{\mu}~V_S \over [ m_\mu~(n-3)]}.
\een
The above derivatives are bounded and it provides that solutions to equations (\ref{rv}) and (\ref{mv})
with these boundary conditions exist locally. From the theory of differential equations \cite{har82}
one concludes that, the local existence theorem requires that left-hand sides of relations (\ref{rv}) and (\ref{mv})
need to be continuous functions of $V$. In other words, the local existence of functions $r_\mu$
and $m_\mu$ is guaranteed even if the density function $\rho(V)$ and pressure function $p(V)$ are not
differentiable at the level surface $V = V_S$.
\par
Let us consider an arbitrary level set $V'_\mu$ in the sense that
the interval $(V'_\mu,~V_S]$ constitutes the maximal interval where the solutions of (\ref{rv})-(\ref{mv}) exist
and pressure is finite $p(V) < \infty$ as well as the following inequalities take place:
\be
r_\mu >0, \qquad r^{n-3}_\mu > 2~m_\mu, \qquad
m_\mu > - {8~\pi \over n - 2}~p~{r^{n-1}_{\mu}},
\ee
then having in mind the relation defining $W_\mu$, one achieves that $W_\mu > 0$, on the considered interval.
As far as ${d r_{\mu} \over d V}$ is concerned, it is greater than zero, so $r_\mu$ is monotonic
and bounded by the relation
\be
R_{\mu} \geq r_\mu >  0.
\ee
The same situation takes place for the other derivative in question, namely for 
${d m_{\mu} \over d V}$.
Because of the fact that ${d m_{\mu} \over d V} > 0$, then it leads to the case that
$m_\mu$ is monotonic and bounded, i.e., we arrive at the relations of the forms
\be
\mu \geq m_\mu > - {8~\pi \over n - 2}~p~{r^{n-1}_{\mu}}.
\ee
Moreover, due to the equations (\ref{rv}) and (\ref{mv}), it can be readily seen that we are left with the following:
\be
{d m_{\mu} \over d r_{\mu}} = {8~\pi \over (n-2) }~\rho~{r^{n-2}_{\mu}}.
\label{dm}
\ee
Next, we assume that $\lim \limits_{V \rightarrow V'_\mu} p = p_\mu < \infty$. By the direct computations
it can be revealed that
\be
{d W_\mu \over dV} + 2~\bigg( n - 2 \bigg)~{W _\mu \over r_\mu}~{d r_\mu \over d V}
- {16~\pi~V \over n - 2}~\bigg[ (n-1)~p + (n-3)~\rho \bigg] = 0,
\label{wmu}
\ee
which constitutes the $n$-dimensional generalization of the equation derived in Ref.\cite{lin94}.
\par
Equation (\ref{wmu}) can be integrated on the interval $(V'_\mu,~V_S]$. One attains the following result:
\be
r^{2(n-2)}_\mu ~W_\mu = \bigg( \mu ~(n-3) \bigg)^2 - \int_{V}^{V_S} { 16~\pi~V~r^{2(n-2)}_\mu \over (n-2)}~
\bigg[ (n-1)~p + (n-3)~\rho \bigg]~dV.
\ee
It can be also revealed that $r^{2(n-2)}_\mu ~W_\mu $ is bounded on the considered interval. The other form of the above 
relation can be provided by
\be
r^{2(n-2)}_\mu ~W_\mu = 
{V^2~\bigg[ {8~\pi \over n - 2}~p~{r^{n-1}_{\mu}} + m_\mu~(n-3) \bigg]^2 \over \bigg(
1 - {2~m_\mu \over r^{n-3}_\mu } \bigg) }.
\label{wmu1}
\ee
The expression $r^{2(n-2)}_\mu ~W_\mu $ will be bounded if the numerator of Eq.(\ref{wmu1}) vanishes in the limit
$V \rightarrow V'_{\mu}$. On its turn, it yields
\be
\lim \limits_{V \rightarrow V'_\mu}~m_\mu~(n-3) = - {8~\pi \over n - 2}~p~{r^{n-1}_{\mu}}.
\ee
If $\lim \limits_{V \rightarrow V'_\mu}~r_\mu > 0$, then we have
\be
\lim \limits_{V \rightarrow V'_\mu}~{2~m_\mu \over r^{n-3}_\mu} = - {16~\pi \over n - 2}~p~{r_{\mu}}^{2}.
\ee
On the other hand, when $\lim \limits_{V \rightarrow V'_\mu}~r_\mu = 0$, we arrive at
the conclusion that $\lim \limits_{V \rightarrow V'_\mu}~{2~m_\mu \over r^{n-3}_\mu} = 0$. Moreover, in other case we get
$\lim \limits_{V \rightarrow V'_\mu}~{2~m_\mu \over r^{n-3}_\mu} \leq 0$. However, this conclusion contradicts our assumption that
$\lim \limits_{V \rightarrow V'_\mu}~{2~m_\mu \over r^{n-3}_\mu} = 1$. Just, it leads us to the conclusion that
\be
sup_{(V'_\mu,~V_S)}~\bigg( {2~m_\mu \over r^{n-3}_\mu} \bigg) < 1.
\ee
Now we shall pay attention to the behaviour of $W_\mu$ function in the limit when $V$ tends to $V'_\mu$. At the beginning 
we consider the case when $\lim \limits_{V \rightarrow V'_\mu}~r_\mu = 0$. In order to find
the limit of $W_\mu$ we arrange the function in question in useful form which implies
\be
W_\mu = V^2~\bigg[
{8~\pi \over n - 2}~p~r_\mu + {m_\mu~(n-3) \over r^{n-2}_\mu} \bigg]^2 ~
\bigg( 1 - {2~m_\mu \over r^{n-3}_\mu } \bigg)^{-1}.
\ee
At first we elaborate the expression
\be
\lim \limits_{V \rightarrow V'_\mu}~\bigg( {m_{\mu}~(n-3) \over r_\mu^{n-2}} \bigg)^2,
\ee
which can be calculated using l'Hospital rule and equation (\ref{dm}). It turns out that the above limit is equal to zero.
Next, we compute
\be
\lim \limits_{V \rightarrow V'_\mu}~\bigg( 1- {2~m_\mu \over r^{n-3}_\mu } \bigg) =
1 - \lim \limits_{V \rightarrow V'_\mu}~\bigg( {2~{d m_\mu \over d r_\mu} \over (n-3)~r^{n-4}_\mu} \bigg) = 1.
\ee
Having all the above in mind, we reach to the conclusion that $\lim \limits_{V \rightarrow V'_\mu}~W_\mu = 0$, which 
in turns 
yields that $V_\mu = V'_\mu$.\\
Secondly, we assume that $\lim \limits_{V \rightarrow V'_\mu}~r_\mu > 0$. As in four-dimensional case \cite{lin94},
$m_\mu$ and $r_\mu$ are absolutely bounded in the considered case. Namely, one has that
\be
sup_{(V'_\mu,~V_S)}~\mid r_\mu \mid = R_\mu, \qquad sup_{(V'_\mu,~V_S)}~\mid m_\mu \mid
< \mu + {8~\pi \over (n-2)}~p~r^{n-1}_\mu.
\ee
Now, we express the relation for ${d r_\mu \over d V}$ in the form as
\be
{d r_\mu \over d V} = {1 \over \sqrt{W_\mu}}~\bigg( 1- {2~m_\mu \over r^{n-3}_\mu } \bigg)^{1 \over 2}.
\ee
Because of the fact that on the interval $(V'_\mu,~V_S]$ the following relation is satisfied
\be
1 > {2~m_\mu \over r^{n-3}_\mu } > - {16~\pi \over (n-2)}~p~r^2_\mu.
\ee
In effect,
the bound of the $r_\mu$ derivative is provided by
\be
\mid {d r_\mu \over d V} \mid < \bigg( {1 + {16~\pi \over (n-2)}~p~R^2_\mu \over W_\mu} \bigg)^{1 \over 2}.
\ee
On the other hand, the bound of the derivative of $m_\mu$ implies
\be
\mid {d m_\mu \over d V} \mid~ <~ \mid {d m_\mu \over d r_\mu} \mid~\mid {d r_\mu \over d V} \mid = 
{8~\pi \over (n-2)}~\rho~R^{n-2}_\mu~
\bigg( {1 + {16~\pi \over (n-2)}~p~R^2_\mu \over W_\mu} \bigg)^{1 \over 2}.
\ee
Just, 
the both derivatives of $r_\mu$ and $m_\mu$ will be bounded at $V = V'_\mu$ if 
$\lim \limits_{V \rightarrow V'_\mu}~W_\mu > 0$. But if it happens, it turns out that the solutions of Eqs.(\ref{rv}) and
(\ref{mv}) can be extended beyond  $V = V'_\mu$ and moreover in this case $r_\mu > 0,~ r^{n-3}_\mu > 2~m_\mu,~
m_\mu~(n-3) > - {8\pi \over (n-2)}~p~r^{n-1}_\mu$, and $p = p_\mu < \infty$ at $V = V'_\mu$.
Using the continuity arguments one can find that these inequalities for $r_\mu,~m_\mu$ can be extended in the vicinity of $V'_\mu$.
But this is not the case, because it violates the assumption that $(V'_\mu,~V_S]$ is the maximal interval on which solutions of 
(\ref{rv}) and (\ref{mv}) can be obtained. It leads to the contradiction and one obtains that 
\be
\lim \limits_{V \rightarrow V'_{\mu}}~W_\mu = 0,
\ee
and then $V'_{\mu} = V_S$. By virtue of the above considerations one can formulate the conclusion as follows:\\
{\bf Theorem:}\\
Let us consider solutions of equations ${d r_{\mu} \over d V}$ and ${d m_{\mu} \over d V}$,~$ r_{\mu}(V)$
and $m_\mu(V)$ which exist on the interval $(V_{\mu},~V_S]$. Suppose that on the interval in question
$p(V)$ is finite and $W_\mu$ is greater than zero. Moreover on the considered interval the following
inequalities hold $r_\mu > ( 2~m_\mu )^{1 \over n-3},$
and $m_{\mu} > - {8~\pi \over n-2}~p~r^{n-1}_\mu.$
Then, if $
\lim \limits_{V \rightarrow V_\mu}~p = p_\mu $ we arrive at the relations
$sup_{(V_\mu,~V_S)}~( {2~m_\mu \over r^{n-3}_\mu}) < 1,$ and 
$\lim \limits_{V \rightarrow V_\mu}~W_\mu = 0.$\\

%%%%%%%%%%%%%%%%%%%%%%%%%%%%%%%%%%%%%%%%%%%%%%%%%%%%%%%%%%%%%%%%%%%%%%%%%%%%%%%%%%%%%%%%%%%%%%%%%%%%%%%%%%%%%%%%%%%%%%%
\vspace{0.1cm}
Furthermore, it happened that the following relation is non-decreasing function of $V$ on the interval $(V_\mu,~V_S]$
\be
m_\mu - {8~\pi \over (n-1)~(n-2)}~\rho~r^{n-1}_\mu.
\ee
It can be readily seen by rearranging (\ref{rv}) and (\ref{mv}) to the form provided by
\be
{d \over d V}~\bigg[
m_\mu - {8~\pi \over (n-1)~(n-2)}~\rho~r^{n-1}_\mu \bigg] = - {8~\pi \over (n-1)~(n-2)}~
\rho~r^{n-1}_\mu ~{d \rho \over d V}.
\label{dww}
\ee
The right-hand side of the above equation is non-negative in the view of (\ref{m3}) and the monotonicity of the equation of state
for the considered case. Further, we can arrange equation containing ${d W_\mu \over d V}$
in the form which yields
\be
{d W_\mu \over d V} - {16~\pi ~V \over (n-1)~(n-2)}
~\bigg[ (n-1)~p + (n-3)~\rho \bigg] =
- {2~(n-3)~(n-2)~V \over r^{n-1}_\mu}~\bigg(
m_\mu - {8~\pi \over (n-1)~(n-2)}~\rho~r^{n-1}_\mu \bigg).
\ee
One achieves that the quantity on the right-hand side, in the brackets, is non-decreasing. Because 
of the fact that $r_\mu$ and $V$
are positive, one draws a conclusion that the right-hand side of the above relation is non-negative 
on the interval in question.\\
Summing it all up, we established the monotonicity of the functions containing contributions of $W_\mu$ and $m_\mu$
for some $V_i > V_\mu$, i.e.,
on the interval $(V_{\mu},~V_i]$. Namely we have the following statement:\\
{\bf Theorem:}\\
Let us consider equation $m_\mu - {8~\pi \over (n-1)~(n-2)}~\rho~r^{n-1}_\mu$ which is a non-decreasing 
function of $V$ on the 
interval $(V_\mu,~V_S]$. If the combination of $W_\mu$ function of the form
${d W_\mu \over d V} - {16~\pi ~V \over (n-1)~(n-2)}~\bigg[ (n-1)~p + (n-3)~\rho \bigg]$ is non-negative 
for some $V_i > V_\mu$,
then it is non-negative for all $V$ belonging to interval $(V_{\mu},~V_i]$.\\
As in four-dimensional case, the solution of Eqs. (\ref{rv}) and (\ref{mv}) for the case when $W_\mu$ 
vanishes, can be split
into two categories. Namely, one can tell about a regular zero of $W_\mu$, when 
$\lim \limits_{V \rightarrow V_\mu} r_\mu = 0$, and an irregular zero of $W_\mu$, when 
$\lim \limits_{V \rightarrow V_\mu} r_\mu > 0$.
In order to determine the additional properties of the zeros of $W_\mu$ function,
let us consider the first case when $\lim \limits_{V \rightarrow V_\mu} W_\mu = 0$ and $\lim 
\limits_{V \rightarrow V_\mu} r_\mu = 0$,
~$\lim \limits_{V \rightarrow V_\mu} p_\mu < \infty $.
We use Eq.(\ref{dww}) and take the limit of both sides of it, when $V \rightarrow V_\mu$. Having in mind
that
\be
\lim \limits_{V \rightarrow V_\mu}~\bigg(
{m_\mu \over r^{n-1}_\mu} \bigg) = \lim \limits_{V \rightarrow V_\mu}~{{d m_\mu \over d r_\mu}
\over (n-1)~r^{n-2}_\mu} = { 8~\pi~\rho \over (n-1)~(n-2)},
\ee
we reach to the conclusion that the following relation is provided:
\be
\lim \limits_{V \rightarrow V_\mu}~\bigg[ {m_\mu \over r^{n-1}_\mu} - { 8~\pi~\rho(V_\mu) \over (n-1)~(n-2)} 
\bigg] = 0.
\ee
Hence it implies that the limit of ${d W_\mu \over d V}$ is given by
\be
\lim \limits_{V \rightarrow V_\mu}~{d W_\mu \over d V} =
{16~\pi~V_\mu~(n-3) \over (n-1)~(n-2)}~\bigg(
\rho(V_\mu) + {n-1 \over n-3}~p_\mu \bigg).
\ee
Now, let us proceed to the case when 
$\lim \limits_{V \rightarrow V_\mu} r_\mu > 0$. In this case we have that 
$\lim \limits_{V \rightarrow V_\mu} m_\mu = - {8~\pi \over (n-2)}~p_\mu~r^{n-1}_\mu$.
Summing it all up, we take the limit of relation (\ref{dww}). It will be provided by
the following expression:
\be
\lim \limits_{V \rightarrow V_\mu}~{d W_\mu \over d V} =
{16~\pi~V_\mu~(n-3) \over (n-2)}~\bigg( \rho(V_\mu) + {1 + (n-2)~(n-3) \over n-3}~p_\mu \bigg).
\ee
%%%%%%%%%%%%%%%%%%%%%%%%%%
Finally, it is worth mentioning that $W_\mu$ implies the second order differential equation
on the interval $(V_\mu,~V_S]$, where one has that $W_\mu$ is greater than zero. The explicit
form of it yields
\ben
{d \over d V}~\bigg[
{1 \over V}~{d W_\mu \over d V} &-& {16~\pi \over n-2}~\bigg[ (n-1)~p + (n-3)~\rho \bigg] \bigg] \\
\nonumber
=
{n-1 \over 2~(n-2)~W_\mu~V}~
\bigg[
{d W_\mu \over d V} &-& {16~\pi ~V \over (n-1)~(n-2)}~\bigg[ (n-1)~p + (n-3)~\rho \bigg] \bigg]~
\times \\ \nonumber
\bigg[
{d W_\mu \over d V} &-& {16~\pi ~V \over (n-1)~(n-2)}~\bigg[ (n-1)~p + (n-3)~\rho \bigg] \bigg],
\een
which can be found by the direct computations.

%%%%%%%%%%%%%%%%%%%%%%%%%%%%%%%%%%%%%%%%%%%%%%%%%%%%%%%%%%%%%%%%%%%%%%%%%%%%%%%%%%%%%%%%%%%%%%%
%%%%%%%%%%%%%%%%%%%%%%%%%%%%%%%%%%%%%%%%%%%%%%%%%%%%%%%%%%%%%%%%%%%%%%%%%%%%%%%%%%%%%%%%%%%%%%%
\section{Conformal factor}
Let us introduce the function $\psi_\mu (V)$, which will constitute a conformal factor for a spatial metric
$g_{ab}$. Namely, $\tg_{ab+} = \Omega_{+}^2~g_{ab}$, where $\Omega_{+} = \psi_{\mu}^{2 \over n-3}$ inside the
considered star and $\psi_{\mu}(V) = \bigg( {1 + V \over 2} \bigg)$ outside the star. 
Having in mind the relation derived in \cite{mas88} and equation(\ref{rv}), it can be revealed that 
inside the star, i.e., $V \in (V_\mu,~V_S]$ this function
satisfies
\be
{d \psi_\mu \over d V} = {(n-3)~\psi_\mu \over 2~\sqrt{W_\mu}~r_\mu}~\bigg(
1 - \sqrt{1 - {2~m_\mu \over r^{n-3}_\mu }} \bigg).
\label{derpsi}
\ee
On the other hand, in the exterior region adjacent to the star, when $V \in [V_S,~1)$ one has 
\be
\psi_{\mu}(V_S) = \bigg( {1 + V_S \over 2} \bigg).
\ee
A tedious calculations reveal that one can get the second order ordinary differential
equation for  $\psi_\mu (V)$. It implies
\be
{d^2 \psi_\mu (V) \over d V^2}
+ {1 \over W_\mu}~\bigg[
- {(n-3) \over 4~(n-2)}~{}^{(g)}R~ \psi_\mu (V) + \bigg(
{d \psi_\mu (V) \over d V}\bigg)~D_aD^a V \bigg] = 0.
\ee
Using relation (\ref{m2}) for the Ricci tensor we attain to the fact that 
${}^{(g)}R = 16~\pi~\rho$. Thus, the above equation for  $\psi_\mu (V)$ reduces to the following:
\be
{d^2 \psi_\mu (V) \over d V^2}
+ {8~\pi \over W_\mu}   ~\bigg[
{V \over (n-3)}~\bigg( (n-3)~\rho + (n-1)~p \bigg)~{d \psi_\mu (V)\over d V}
- {(n-3) \over 2~(n-2)}~\rho~\psi_\mu (V) \bigg] = 0.
\label{pss}
\ee
Our main aim will be to find the sign of the function ${d^2 \psi_\mu (V) \over d V^2}$. Let us define
the function $f_\mu$ in the form as
\be
f_\mu = {4~\pi~(n-3) \over n - 2}~\rho - {8 \pi~V \over n-3}~\bigg( (n-3)~\rho + (n-1)~p \bigg)~{1 \over
\psi_{\mu}}~{d \psi_\mu \over d V}.
\ee
Next we calculated ${d f_\mu \over d V}$. Relations (\ref{pss}) and (\ref{m3}) clearly imply that
\ben \label{fmu}
8 \pi~V~\bigg( (n-3)~\rho &+& (n-1)~p \bigg)~{d f_\mu \over d V}
= {16~\pi^2~(n-3) \over (n-2)^2}~\bigg[
\bigg( (n-3) + 4~(n-2) \bigg)~\rho^2 \\ \nonumber
&-& 2~(n-1)(n-2)~p~(\rho + p)~\kappa \bigg] \\ \nonumber
&+& (n-3)~f^2_\mu - f_\mu~\bigg[
8~\pi~(n-3)~(\rho + p)~\kappa + {64~\pi^2~V^2 \over (n-3)~W_\mu}~\bigg( (n-3)~\rho + (n-1)~p \bigg)^2 \\ \nonumber
&-& {8~\pi~\rho \over n-2}~~\bigg( (n-3)~\rho + (n-1)~p \bigg) \bigg],
\een
where $\kappa = {d p \over d \rho}.$
One can observe that the function $f_\mu$ was chosen in such a way that ${d^2 \psi_\mu (V) \over d V^2} = \alpha~f_\mu$,
where $\alpha$ is a constant value. Thus, the sign of $f_\mu$ will be crucial to establish the sign of the second
derivative with respect to $V$ from $\psi_\mu$. If this equation implies that $f_\mu < 0$ for some $(V_\mu,~V_S]$, we get
that ${d f_\mu \over d V} > 0$, under the following condition:
\be
\bigg( (n-3) + 4~(n-2) \bigg)~\rho^2 - 2~(n-1)(n-2)~p~(\rho + p)~\kappa \geq 0.
\label{cond}
\ee
But $\lim \limits_{V \rightarrow V_\mu} f_\mu(V) = 0$, when $V_\mu$ is a regular zero and
$\lim \limits_{V \rightarrow V_\mu} f_\mu(V) = \infty$ if one has the case of irregular zero for $V_\mu$.
The last statement yields that $f_\mu$ is not negative at $V = V_\mu$ and it is impossible
to be negative for larger values of $V$ unless ${d f_\mu \over d V} < 0$ there. In turn, this is
in contradiction to the conclusion from the inspection of equation (\ref{fmu}). Just, we have that
$f_\mu \geq 0$, which in turn implies that ${d^2 \psi_\mu \over d V^2} \geq 0$.\\
Thus, we arrive at the following conclusion:\\
{\bf Theorem:}\\
Let us assume that $\rho = \rho(p)$ is nonnegative, not decreasing function of pressure, satisfying the
additional condition (\ref{cond}). It turns out that the function $\psi_\mu$ defined by the relation
$\psi_\mu = 1/2~(1 + V)$ and its derivative with respect to $V$ defined by (\ref{derpsi})
provides that ${d^2 \psi_\mu \over d V^2} \geq 0$.\\

%%%%%%%%%%%%%%%%%%%%%%%%%%%%%%%%%%%%%%%%%%%%%%%%%%%%%%%%%%%%%%%%%%%%%%%%%%%%%%%%%%%%%%%%%%%%%%%%%%%%%%%
%%%%%%%%%%%%%%%%%%%%%%%%%%%%%%%%%%%%%%%%%%%%%%%%%%%%%%%%%%%%%%%%%%%%%%%%%%%%%%%%%%%%%%%%%%%%%%%%%%%%%%
\section{Nonexistence of perfect fluid matter in static n-dimensional black hole background}
The basic idea in our treatment of the problem in question will be to use the positive mass theorem 
in Bartnik formulation \cite{pos}. 
As in the proofs of the uniqueness of static $n$-dimensional black holes, we assume the validity
of the positive mass theorem which states
that if one considers asymptotically flat, complete,
orientable Riemanian manifold with non-negative scalar Ricci curvature and vanishing mass, then the
considered manifold is isometric to $(E^{n-1},~\delta_{ij})$ manifold.\\
To begin with, let us consider two conformal transformations provided by 
\be
\tg_{ab_\pm} = \Omega_{\pm}^2~g_{ab},
\ee
where $ \Omega_{+}$ is given by relations from the later section 
and $ \Omega_{-} = \xi_{\mu }^{2 \over n-3}$, where $ ~\xi_{\mu}
= {1 \over 2} (1-V)$. Hence we have two manifolds $(\Sigma_{\pm},~\tg_{ij \pm})$. 
Because of this fact we take into account two 
copies of the hypersurface $\Sigma_{+}$ and $\Sigma_{-}$ and define metric 
on them. By $\p_{nodeg}\Sigma$ one denotes all the components of the boundary of 
$\Sigma$ which correspond to non-degenerate components of the black hole event
horizons. One gets the following relations:
\ben
\Sigma_{+} &=& \Sigma, \qquad \tg_{ij+},\\
\Sigma_{-} &=& \Sigma \cup \{ p_{i} \}, \qquad \tg_{ij-},
\een
where $\Sigma \cup \{ p_{i} \}$ describes a one point compactification of all asymptotically flat regions
of the hypersurface $\Sigma$. Moreover, $p_i$ denotes a point of the adequate asymptotically flat region.
\par
It should be 
pointed out that the positive mass theorem \cite{pos} can not be implemented for 
$(\Sigma_{\pm},~\tg_{ij \pm})$. In order to fulfill the requirements of the aforementioned theorem
we have to have the Riemanian manifold composed of two copies of $\Sigma_{\pm}$.
We obtain $\hat \Sigma = \Sigma_{+} \cup \Sigma_{+} \cup \p_{nodeg}\Sigma$. The differential
structure on $\hat \Sigma$ is provided by gluing $\bar \Sigma_{+} = \Sigma_{+}
\cup \p_{nodeg}\Sigma$ with $\bar \Sigma_{-} = \Sigma_{-}
\cup \p_{nodeg}\Sigma$, and identifying $\p_{nodeg}\Sigma$ considered as a subset of $\bar \Sigma_{+}$ with
$\p_{nodeg}\Sigma$ considered as a subset of $ \bar \Sigma_{-}$, using the identity map. 
The metric tensor defined on $\Sigma_{+} \cup \Sigma_{-}$ can be extended by
continuity to smooth metric on $\hat \Sigma$. It implies
\be
{\hat g}_{ij}\mid_{\Sigma_{+}} = \tg_{ij +}  \qquad 
{\hat g}_{ij}\mid_{\Sigma_{-}} = \tg_{ij -}.
\ee
In the case when $\p_{nodeg}\Sigma = 0$, we arrive at ${\hat \Sigma} = \Sigma$,~
$\hat g_{ij} = \tg_{ij}$.\\
Then, on the hypersurface $\Sigma_{+}$
we have the asymptotical behaviour of the metric tensor $\tg_{ij+}$ of the form
\be
\tg_{ij+} = \delta_{ij} + \cO \bigg({1 \over r^{n-2}}\bigg),
\label{asp}
\ee
while on $\Sigma_{-}$ hypersurface, the asymptotic behaviour of the metric $\tg_{ij-}$ yields
\be
\tg_{ij-} = \bigg( {m_\mu \over 2~r^{n-3}} \bigg)^{4 \over n-3} 
g_{ij} + \cO \bigg( {1 \over  r^{2n -3}} \bigg).
\label{asm}
\ee
Let us examine the conformally rescaled Ricci curvature scalar $\tR$ on both
hypersurfaces in question. In the case of 
$\Sigma_{-}$ hypersurface we get the following expression:
\be
\tR(\Omega^2_{-}g_{ab}) = 
8~\pi~\bigg( {1 - V \over 2} \bigg)^{- {n+1 \over n-3}}~\bigg[ \rho~\bigg( V + 1 \bigg) + 2 \bigg(
{n - 1 \over n - 3} \bigg)~V~p \bigg].
\ee
On the other hand, on
$\Sigma_{+}$ hypersurface, $\tR(\Omega^2_{+}g_{ab})$ may be written in the form as
\be
\tR(\Omega^2_{+}g_{ab}) = \bigg( \tW - W \bigg)~{4~(n-2) \over n-3}~
{1 \over \psi^{n+1 \over n-3}_\mu}~{d^2 \psi_\mu \over d V^2},
\ee
where we have denoted $W = D_iV~D^i V$. 
We remark that having in mind the asymptotical value of $V$, the asymptotic
behaviour of $W$ can be deduced. It is given by the expression
\be
W = {(n-3)^2~m^2_\mu \over r^{2(n-2)}} + {\cal O}\bigg( {1 \over  r^{2n-3}} \bigg).
\ee
Because of the fact that ${d^2 \psi_\mu \over d V^2} \geq 0$, in order to have the non-negative
conformally rescaled Ricci scalar tensor we assume that $\bigg( \tW - W \bigg) \geq 0$,
where $\tW = W_\mu$.\\
%%%%%%%%%%%%%%%%%%%%%%%%%%%%%%%%%%%%%%%%%%%%%%%%%%%%%%%%%%%%%%%%%%%%%%%%%%%%%%%%%%%%%%%%5
\par
Let us give some remarks concerning the above inequality, which will be crucial in what follows.
In four-dimensional case \cite{lin94} the proof 
of the aforementioned inequality is based on the identity due to Lindblom \cite{lin80}
which consists of the square of the Bach-Cotton tensor. It is well known that on three-dimensional
manifold the Weyl tensor vanishes and the conformal properties of three-dimensional manifold are described 
by the Bach-Cotton tensor.
In $n$-dimensional case one can generalize this idea and use the Bach-Cotton tensor in 
$(n-1)$-dimensional spacetime. Then
the boundary point maximum principle to prove the inequality can be implemented \cite{gil01}. 
\par
On this account it yields \cite{rog}
\ben
{}^{(g)}C_{ijk}~{}^{(g)}C^{ijk} &=& 
\alpha~{W ~\na_k \na^k W \over 2~V^4} - {8~\pi~\alpha W \over V^3}~\bigg(
{2n-5 \over n-2} \bigg)~D_i \rho~D^i V + {16~\pi ~\alpha~W^2~\rho \over (n-2)~V^4} \\ \nonumber
&+&
{8~\pi~(n-2) \over V^3~(n-3)^2}~D_iV ~D^i W~\bigg[ (n-2)~p + (n-3)~\rho \bigg]
- {(n-2) \over 4~V^4~(n-3)^2}~D_a W~D^a W \\ \nonumber
&+& {64~\pi^2~W \over V^2~(n-2)^2~(n-3)^2}~\bigg[
\alpha_1~p^2 + \alpha_2~p~\rho + \alpha_3~\rho^2 \bigg],
\een
where $\alpha,~\alpha_1,~\alpha_2,~\alpha_3$ are constant coefficients \cite{rog}.
Next, the maximum principle can be used in the proof. What is more, one can also investigate 
the identity including the Weyl tensor
describing the conformal properties in $n > 3$ dimensions. After tedious computations it can be revealed that 
the square of the $(n-1)$-dimensional Weyl tensor implies
\ben
{}^{(g)}C_{ijkl}~{}^{(g)}C^{ijkl} &=&
\bigg( {n-2 \over n-3} \bigg)^2~\bigg[
64~\pi^2 ~p^2~(n-1) - 128~\pi^2~p~\bigg[ (n-1)~p + (n-3)~\rho \bigg]
\\ \nonumber
&+& {64~\pi^2~(n-3) \over (n-2)^2}~\bigg[ (n-1)~p + (n-3)~\rho \bigg]^2 \bigg]\\ \nonumber
&-&
{4 \over n-2}~\bigg[
{64~\pi^2~(n-1) \over n-2}~(\rho - p)^2 + {128~\pi^2 \over (n-2)^2}
~(\rho - p)~\bigg[ (n-1)~p + (n-3)~\rho \bigg] \bigg] 
\\ \nonumber
&+&
{8~(n-1)\over (n-2)~(n-3)^2}~256~\pi^2\rho^2 \\ \nonumber
&-& {4~(n-2)^3 \over V^2~(n-3)^2~(n-2)}~\bigg[
{1 \over 2}~\na_k \na^k W + {16~\pi \over n-2}~W~\rho
- \bigg( {2n -5 \over n-2} \bigg)~D_i \rho~D^i V \bigg]
\een
In accordance with the above, 
the maximum principle should be used to prove the inequality in question.
\par
%%%%%%%%%%%%%%%%%%%%%%%%%%%%%%%%%%%%%%%%%%%%%%%%%%%%%%%%%%%%%%%%%%%%%%%%%%%%%%%%%%%%%%%%%%%%%%%%%%%%%%%
Returning to the problem in question,
one has that both Ricci scalar curvature tensors are non-negative and moreover
equation (\ref{asp}) implies that the total ADM mass also vanishes. As a consequence of the positive mass
theorem the hypersurface $\hat \Sigma$ is isometric to flat manifold
or in the other words, the Cauchy surface in question is conformally flat. Consequently, we have that
\be
\tR(\Omega^2_{-}g_{ab}) = \tR(\Omega^2_{+}g_{ab}) = 0.
\ee
The above relation provides the following:
\be
W = W_\mu, \qquad \rho = p = 0.
\ee
The second condition yields that the spacetime under consideration ought to be empty, i.e.,
one excludes any static configuration composed of $n$-dimensional black hole and a perfect fluid star.\\
Finally, the 
above considerations enable us to formulate the main conclusion
of our work.\\
{\bf Theorem:}\\
Let us consider a static black hole spacetime with an asymptotically timelike Killing vector field
$k_\mu$. Assume that one considers in such background a perfect fluid star which has the surface of the
level surface set $\{V =V_S>0 \}$. Suppose moreover that equation of state $p = p(\rho)$ is provided by
$$
\bigg( (n-3) + 4~(n-2) \bigg)~\rho^2 - 2~(n-1)(n-2)~p~(\rho + p)~\kappa \geq 0, $$
and the inequality $(W - W_\mu)_{V=0} \leq 0$, where $W_\mu$ is defined by relation (\ref{wdef}) and $W = D_jV~D^jV$,
is satisfied on the black hole event horizon. Then under the above conditions,
a perfect fluid star cannot exist in a static $n$-dimensional black hole spacetime.\\ 

%%%%%%%%%%%%%%%%%%%%%%%%%%%%%%%%%%%%%%%%%%%%%%%%%%%%%%%%%%%%%%%%%%%%%%%%%%%%%%%
\par
Now, we proceed to give some remarks concerning spherical symmetry of the general static relativistic
$n$-dimensional perfect fluid stellar model. The arguments that lead us to state that
the spatial geometry $g_{ab}$ of a static stellar model described by equations
(\ref{m1})-(\ref{m3}) has spherical symmetry are mainly the same as in the uniqueness proof
of static $n$-dimensional black holes \cite{gib}. We outline the crucial steps of them.\\
To begin with, one can choose $V$ as a local coordinate in the neighborhood $U \in \Sigma$.
With this understanding the metric on the hypersurface $\Sigma$ can be written in the form as
\be
ds^2 = \eta^2~dV^2 + h_{AB}~dx^A~dx^B,
\ee
where we set $\eta^2 =( D_m V~D^mV)^2$. The $x_A$-coordinates are chosen in such a way that their trajectories
are orthogonal to each of the level set.  One can also rewrite $g_{ij}$ in a 
conformally flat form \cite{gib}
\be
g_{ij} = {\cal U}^{1 \over n-3} \delta_{ij},
\label{gg}
\ee
where we have defined a smooth function on $(\hat \Sigma, \delta_{ij})$,
namely ${\cal U} = {2 \over 1 + V }$. It can be easily found that
Einstein equations of motion reduces
to the Laplace equation on the $(n - 1)$ Euclidean manifold 
$
\na_{i}\na^{i}{\cal U} = 0,
$
where $\na$ is the connection on a flat manifold. Accordingly, we can adopt for the metric $\delta_{ij}$
in the flat base space the following metric:
\be 
\delta_{ij} dx^{i}dx^{j} = \teta^{2} d{\cal U}^2 + {\tilde h}_{AB}dx^{A}dx^{B}.
\ee
Further, 
one can show that the embedding of the stellar surface into the Euclidean
$(n-1)$-dimensional space is totally umbilical \cite{kob69}. This embedding have to be 
hyperspherical, i.e., each of the connected components of the surface in question
is a geometric sphere with a certain radius determined by the value of
$\eta \mid_{surf}$.
One can always locate one of the 
connected component of the stellar surface at $\eta = \eta_{0}$ without loss of generality.
It turns out that, 
we have to do with a boundary value problem for the Laplace equation 
on the base space $\Omega = E^{n-1}/B^{n-1}$ with the
Dirichlet boundary condition ${\cal U} \mid_{surf}$ and the asymptotic
decay condition ${\cal U} = 1 + {\cal O} \bigg( {1 \over r^{n-3}} \bigg)$.

%%%%%%%%%%%%%%%%%%%%%%%%%%%%%%%%%%%%%%%%%%%%%%%%%%%%%%%%%%%%%%%%%%%%%%%%%%%%%%%%%%%%%%%%%%%%%%%%
%%%%%%%%%%%%%%%%%%%%%%%%%%%%%%%%%%%%%%%%%%%%%%%%%%%%%%%%%%%%%%%%%%%%%%%%%%%%%%%%%%%%%%%%%%%%%%%%
\section{Inequality}
In this section we restrict our attention to the physical meaning of
our hypothesis. On this account, let us 
consider the case when $(W - W_\mu)_{V = 0} \leq 0$. To begin with,         
we compute the surface gravity $\kappa$ of $n$-dimensional static black hole. 
It yields the following relation:
\be
\kappa = \sqrt{- {1 \over 4}~g^{tt}~g^{rr} (g_{tt,r})} \mid_{r=r_{BH}} =
{(n-3)~m^{3-n}_{BH} \over 2^{n-2}}.
\ee
Then, letting $V=0$ and having in mind the relation for $W_\mu$, we receive
\be
\sqrt{W_{V=0}} = {n-3 \over 2^{{n-2 \over n-3}}~\mu^{1 \over n-3}}.
\ee
The Killing vectors for static asymptotically flat spacetime may be used to find the coordinate independent
expression for black hole mass. By virtue of it, we can readily verify that the following
expression is satisfied: 
\be
\mu^{1 \over n-3} \leq { 1 \over \bigg(
{2^{n-2 \over n-3}}~\kappa \bigg)^{1 \over n-3} }
= \bigg[
{(n-2) ~A_{BH} \over 8~\pi~2^{n-2}~M_{BH}} \bigg]^{1 \over (n-3)^2}.
\ee
Moreover, having in mind that \cite{mye86} 
\be
{n-3 \over n-2}~M_{BH} = {\kappa~A_{BH} \over 8~\pi},
\ee
one arrives at the relation provided by
\be
\mu^{1 \over n-3} \leq  (m_{BH})^{1 \over n-3},
\ee
where black hole mass yields
\be
M_{BH} = {(n-2) \over 8~\pi}~\Omega_{(n-2)}~m_{BH}.
\ee
In order to establish the upper limit on $m_\mu$, we consider the next inequality, which takes place
on the surface of the perfect fluid star $(W - W_\mu)_{V=V_S} \leq 0$. In the case 
under consideration we get
\be
\sqrt{W_{\mu}} = {(n-3)~(1 -V_S)^{n-2 \over n-3} \over 2^{n-2 \over n-3}~\mu^{1 \over n-3}}.
\ee
Let us take the asymptotic value of $\sqrt{W_{\mu}}$, on the left-hand side of the above equation
\be
m_\mu \leq {\bigg( 1 - V^2_S \bigg)^{n-2 \over n-3} \over 2^{n-2 \over n-3}~\mu^{1 \over n-3}}
~r^{n-2}_\mu.
\ee
Integrating over the $(n-2)$-dimensional sphere we get the expression of the form
\be
\mu^{1 \over n-3} \leq
{\bigg( 1 - V^2_S \bigg)^{n-2 \over n-3}~A_{star} \over 8~\pi~2^{n-2 \over n-3}~M_{star}},
\ee
where $A_{star} = \Omega_{(n-2)}~r^{n-2}_S$ and $M_{star} = {n-2 \over 8~\pi}~\Omega_{(n-2)}~m_\mu$,
are the area of the star surface and its mass, respectively.
Hence, we can readily write down 
\be
\mu^{1 \over n-3} \leq  (m_{\mu})^{1 \over n-3}.
\ee
On this account, it is customary to conclude that
\be
(m_{\mu})^{1 \over n-3} \leq (m_{BH})^{1 \over n-3}.
\label{ineq}
\ee
Taking into account the inequality (\ref{ineq}) we can assert that our main theorem states that a perfect
fluid star of a smaller mass than a black hole mass in not able to exist outside a static black hole spacetime.           

%%%%%%%%%%%%%%%%%%%%%%%%%%%%%%%%%%%%%%%%%%%%%%%%%%%%%%%%%%%%%%%%%%%%%%%%%%%%%%%%%%%%%%%%%%%%%%%%%%%%%%%%%%%%%%%%%%
%%%%%%%%%%%%%%%%%%%%%%%%%%%%%%%%%%%%%%%%%%%%%%%%%%%%%%%%%%%%%%%%%%%%%%%%%%%%%%%%%%%%%%%%%%%%%%%%%%%%%%%%%%%%%%%%%
\section{Conclusions}
In our paper we have considered matter fields of perfect fluid in the spacetime of
$n$-dimensional static asymptotically flat black hole. The main aim was to prove the
uniqueness of such a configuration. As to how this might be done, can be inferred from
the generalization of the method using the positive mass theorem \cite{pos} for showing
the uniqueness of static black holes. Namely, considering the conformal transformation
on the hypersurface $t = const$, one finds that the Ricci scalar curvature tensor is non-negative
and moreover $\Sigma$ becomes a hypersurface with zero ADM mass. As a consequence, the manifold
in question satisfies the conditions to apply the positive mass theorem, which in turn
yields that $\Sigma$ is isometric to flat manifold. Then, the inspection of the conformally
rescaled Ricci scalar tensors enables one to find that $\rho = p = 0$. It means that we
excluded any static configurations composed of $n$-dimensional black hole and a perfect fluid
star. The aforementioned proof was achieved under the auxiliary inequality, which means that
the mass of the perfect fluid star is smaller than the mass of static $n$-dimensional
black hole.

%%%%%%%%%%%%%%%%%%%%%%%%%%%%%%%%%%%%%%%%%%%%%%%%%%%%%%%%%%%%%%%%%%%%%%%%%%%%%%%%%%%%%%%%%%%%%%%%
%%%%%%%%%%%%%%%%%%%%%%%%%%%%%%%%%%%%%%%%%%%%%%%%%%%%%%%%%%%%%%%%%%%%%%%%%%%%%%%%%%%%%%%%%%%%%%%%
%\begin{appendix}

%\section{Irred   } 
%\label{irtf}
%\end{appendix}
\begin{acknowledgments}
MR was partially supported by the grant of the National Science Center
$2011/01/B/ST2/00408$.
\end{acknowledgments}

%%%%%%%%%%%%%%%%%%%%%%%%%%%%%%%%%%%%%%%%%%%%%%%%%%%%%%%%%%%%%%%%%%%%%%%%%%%%%%%%%%%%%%%%%%%%%%%%%%%%%%%
%%%%%%%%%%%%%%%%%%%%%%%%%%%%%%%%%%%%%%%%%%%%%%%%%%%%%%%%%%%%%%%%%%%
%%%%%%%%%%%%%%%%%%%%%%%%%%%%%%%%%%%%%%%%%%%%%%%%%%%%%%%%%%%%%%%%%%%%%%%%%%%%%%%%%
%%%%%%%%%%%%%%%%%%%%%%%%%%%%%%%%%%%%%%%%%%%%%%%%%%%%%%%%%%%%%%%%%%%%%%%%%%%%%%%%%


\begin{thebibliography}{99}
%
\def\cmp#1#2#3{{ Commun. Math. Phys.} {\bf #1}, #2 (#3)}
\def\lmp#1#2#3{{ Lett. Math. Phys.} {\bf #1}, #2 (#3)}
\def\cpam#1#2#3{{ Commun. Pure Appl. Math.} {\bf #1}, #2 (#3)}
\def\hpa#1#2#3{{ Hell. Phys. Acta} {\bf #1}, #2 (#3)}
\def\grg#1#2#3{{ Gen. Rel. Grav.} {\bf #1}, #2 (#3)}
\def\pr#1#2#3{{ Phys. Rev.} {\bf #1}, #2 (#3)}
\def\prl#1#2#3{{ Phys. Rev. Lett.} {\bf #1}, #2 (#3)}
\def\prd#1#2#3{{ Phys. Rev. D} {\bf #1}, #2 (#3)}
\def\pl#1#2#3{{ Phys. Lett} {\bf #1}, #2 (#3)}
\def\pla#1#2#3{{ Phys. Lett. A} {\bf #1}, #2 (#3)}
\def\plb#1#2#3{{ Phys. Lett. B} {\bf #1}, #2 (#3)}
\def\prep#1#2#3{{ Phys. Reports} {\bf #1}, #2 (#3)}
\def\phys#1#2#3{{ Physica} {\bf #1}, #2 (#3)}
\def\jcp#1#2#3{{ J. Comput. Phys.} {\bf #1}, #2 (#3)}
\def\jmp#1#2#3{{ J. Math. Phys.} {\bf #1}, #2 (#3)}
\def\jpm#1#2#3{{ J. Phys. A: Math. Gen.} {\bf #1}, #2 (#3)}
\def\cpr#1#2#3{{ Computer Phys. Rept.} {\bf #1}, #2 (#3)}
\def\cqg#1#2#3{{ Class. Quantum Grav.} {\bf #1}, #2 (#3)}
\def\cma#1#2#3{{ Computers Math. Applic.} {\bf #1}, #2 (#3)}
\def\mc#1#2#3{{ Math. Compt.} {\bf #1}, #2 (#3)}
\def\apj#1#2#3{{ Astrophys. J.} {\bf #1}, #2 (#3)}
\def\apjs#1#2#3{{ Astrophys. J. Suppl.} {\bf #1}, #2 (#3)}
\def\acta#1#2#3{{ Acta Astronomica} {\bf #1}, #2 (#3)}
%%%%%%%%%%%%%%%%%%%%%%%%%%%%%%%%%%%%%%%%%%%%%%%%%%%%%%%%%%%%%%%%%%%%%%%%%%
\def\apl#1#2#3{{Ann. Physik. (Leipzig)} {\bf #1}, #2 (#3)}
\def\anp#1#2#3{{Ann. Phys. } {\bf #1}, #2 (#3)}
\def\sa#1#2#3{{ Sov. Astro.} {\bf #1}, #2 (#3)}
\def\sia#1#2#3{{ SIAM J. Sci. Statist. Comput.} {\bf #1}, #2 (#3)}
\def\aa#1#2#3{{ Astron. Astrophys.} {\bf #1}, #2 (#3)}
\def\mnras#1#2#3{{ Mon. Not. R. astr. Soc.} {\bf #1}, #2 (#3)}
\def\npb#1#2#3{{ Nucl. Phys. B} {\bf #1}, #2 (#3)}
\def\prsla#1#2#3{{ Proc. R. Soc. London, Ser. A} {\bf #1}, #2 (#3)}
\def\jhep#1#2#3{{ JHEP} {\bf #1}, #2 (#3)}
\def\nuc#1#2#3{{Nuovo Cimento B } {\bf #1}, #2 (#3)}
\def\ijmp#1#2#3{{Int. J. Mod. Phys. D} {\bf #1}, #2 (#3)}
\def\atmp#1#2#3{{Adv. Theor. Math. Phys.} {\bf #1}, #2 (#3)}
\def\ptps#1#2#3{{Prog. Theor. Phys. Suppl.} {\bf #1}, #2 (#3)}
\def\lmp#1#2#3{{Lett. Math. Phys. } {\bf #1}, #2 (#3)}
\def\mmj#1#2#3{{Mich. Math. j. } {\bf #1}, #2 (#3)}
\def\hepph#1#2{{ hep-ph }{\bf #1} (#2)}
\def\hepth#1#2{{ hep-th }{\bf #1} (#2)}
\def\grqc#1#2{{ gr-qc }{\bf #1} (#2)}
\def\ibid#1#2#3{{ {\it ibid.} }{\bf #1}, #2 (#3)}
%
%%%%%%%%%%%%%%%%%%%%%%%%%%%%%%%%%%%%%%%%%%%%%%%%%%%%%%%%%%%%%%%%%%%%%%
\bibitem{isr67}
W.Israel, \pr{164}{1776}{1967}.
\bibitem{isr68}
W.Israel, \cmp{8}{245}{1968}.
\bibitem{mil73}
H.M\"uller zum Hagen, C.D.Robinson, and H.J.Seifert, \grg{4}{53}{1973},\\
H.M\"uller zum Hagen, C.D.Robinson, and H.J.Seifert, \grg{5}{61}{1974}
\bibitem{rob77}
C.D.Robinson, \grg{8}{65}{1977}. 
\bibitem{bun87}
G.L.Bunting and A.K.M.Masood-ul-Alam, \grg{19}{147}{1987}.
\bibitem{rub88}
P.Ruback, \cqg{5}{L155}{1988}. 
\bibitem{mas92}
A.K.M.Masood-ul-Alam, \cqg{9}{L53}{1992}.
\bibitem{heu94}
M.Heusler, \cqg{11}{L49}{1994}.

\bibitem{chr99a}
P.T.Chru\'sciel, \cqg{16}{661}{1999}.
\bibitem{chr99b}
P.T.Chru\'sciel, \cqg{16}{689}{1999}.
\bibitem{chr07}
P.T.Chru\'sciel and P.Tod, \cmp{271}{577}{2007}.


\bibitem{car73}
B.Carter,  in {\it Black Holes}, edited
by DeWitt C and DeWitt B.S (Gordon and Breach, New York) (1973).
\bibitem{car87}
B.Carter,  in {\it Gravitation and Astrophysics}, edited
by B.Carter and J.B.Hartle (Plenum Press, New York) (1987).
\bibitem{rob75}
C.D.Robinson, \prl{34}{905}{1975}.

\bibitem{maz82}
P.O.Mazur, \jpm{15}{3173}{1982},\\
P.O.Mazur, {\it Black Hole Uniqueness Theorems}
{\it Preprint }, \hepth{0101012}{2001}.
\bibitem{bun}
G.L.Bunting, PHD thesis, Univ.of New 
England, Armidale N.S.W. (1983).
\bibitem{car85}
B.Carter, \cmp{99}{565}{1985}.

\bibitem{heub}
M.Heusler, {\it Black Hole Uniqueness Theorems} 
(Cambridge: Cambridge University Press) (1997).


\bibitem{lowen}
A.K.M.Masood-ul-Alam, \cqg{10}{2649}{1993},\\ 
M.G\"urses and E.Sermutlu, \ibid{12}{2799}{1995},\\
M.Rogatko, \ibid{14}{2425}{1997},\\
M.Rogatko, \ibid{58}{044011}{1998},\\
M.Rogatko, \ibid{59}{104010}{1999},\\
M.Rogatko, \ibid{82}{044017}{2010},\\
M.Rogatko, \cqg{19}{875}{2002},\\
S.Tomizawa, Y.Yasui, and A.Ishibashi, \prd{79}{124023}{2009},\\
S.Tomizawa, Y.Yasui, and A.Ishibashi, \ibid{81}{084037}{2010},\\
J.B.Gutowski, \jhep{0408}{049}{2004},\\
J.P.Gauntlett, J.B.Gutowski, C.M.Hull, S.Pakis, and H.S.Real, \cqg{20}{4587}{2003},\\
M.Mars and W.Simon, \atmp{6}{279}{2003}.

\bibitem{gib}
G.W.Gibbons, D.Ida, and T.Shiromizu, \ptps{148}{284}{2003},\\
G.W.Gibbons, D.Ida, and T.Shiromizu, \prd{66}{044010}{2002},\\
G.W.Gibbons, D.Ida, and T.Shiromizu, \prl{89}{041101}{2002},\\
M.Rogatko, \cqg{19}{L151}{2002}.

\bibitem{rog0306}
M.Rogatko, \prd{67}{084025}{2003},\\
M.Rogatko, \ibid{73}{124027}{2006}.


\bibitem{rog04}
M.Rogatko, \prd{70}{044023}{2004}.
\bibitem{rog05}
M.Rogatko, \prd{71}{024031}{2005}.

\bibitem{mye86}
R.C.Myers and M.J.Perry, \anp{172}{304}{1986}.

\bibitem{emp02}
R.Emparan and H.S.Reall, \prl{88}{101101}{2002}.

\bibitem{haw73}
S.W.Hawking and G.F.R.Ellis, {\it The Large Scale Structure of Space-Time} (Cambridge University
Press, Cambridge, England, 1973).

\bibitem{gal05}
G.J.Galloway and R.Schoen, \cmp{266}{571}{2006}.

\bibitem{hel05}
C.Helfgott, Y.Oz, and Y.Yanay, \jhep{02}{025}{2005}.

\bibitem{nrot}
Y.Morisawa and D.Ida, \prd{69}{124005}{2004},\\
M.Rogatko, \ibid{70}{084025}{2004},\\
Y.Morisawa, S.Tomizawa, and Y.Yasui, \ibid{77}{064019}{2008},\\
M.Rogatko, \ibid{77}{124037}{2008},\\
S.Hollands, A.Ishibashi, and R.M.Wald, \cmp{271}{699}{2007},\\
S.Hollands and S.Yazadjiev, \ibid{283}{749}{2008},\\
S.Hollands and S.Yazadjiev, \cqg{25}{095010}{2008},\\
D.Ida, A.Ishibashi, and T.Shiromizu, \ptps{189}{52}{2011}.


\bibitem{haj74}
P.Hajicek, \jmp{15}{1554}{1974},\\
G.W.Gibbons, \prd{15}{3530}{1974}.
                  
\bibitem{loh84}
D.Lohiya, \prd{30}{1194}{1984}.

\bibitem{smo}
F.Finster, J.Smoller, and S.T.Yau, \atmp{4}{1231}{2000}.

\bibitem{nak12}
L.Nakonieczny and M.Rogatko, \prd{85}{124050}{2012}.

\bibitem{mas87}
A.K.M.Masood-ul-Alam, \cqg{4}{625}{1987}.
\bibitem{mas88}
A.K.M.Masood-ul-Alam, \cqg{5}{409}{1988}.
\bibitem{lin88}
L.Lindblom, \jmp{29}{436}{1988}.
\bibitem{bei92}
R.Beig and W.Simon, \cmp{144}{373}{1992}.
\bibitem{lin80}
L.Lindblom, \jmp{21}{1455}{1980}.
\bibitem{lin94}
L.Lindblom and A.K.M.Masood-ul-Alam, \cmp{162}{123}{1994}. 

\bibitem{shi06}
T.Shiromizu and S.Yamada, \jmp{47}{112502}{2006}.

\bibitem{gil01}
D.Gilbarg and N.S.Trudinger, {\it Elliptic Partial Differential Equations of Second Order},
(Berlin,~Heidelberg,~New York 2001).
\bibitem{har82}               
P.Hartman, {\it Ordinary Differential Equations} (Boston, Birkh\"auser 1982).


\bibitem{pos}
R.Bartnik, \cpam{39}{661}{1986},\\
R.Schoen and S-T.Yau, \cmp{65}{45}{1979},\\
E.Witten, \cmp{80}{381}{1981}.



\bibitem{rog}
M.Rogatko to be published.

     
\bibitem{kob69}
S.Kobayashi and K.Nomizu, {\it Foundations of Differential Geometry}, (New York, Interscience, 1969). 


%%%%%%%%%%%%%%%%%%%%%%%%%%%%%%%%%%%%%%%%%%%%%%%%%%%%%%%%%%%%%%%%%%%%%%%%%%%%%%%%%%%%%%%%%%%%%%
\end{thebibliography}
\end{document}